\documentclass[
aps,%
12pt,%
final,%
notitlepage,%
oneside,%
onecolumn,%
nobibnotes,%
nofootinbib,%
superscriptaddress,%
noshowpacs,%
centertags]%
{revtex4}

\begin{document}

\title{ANALYSIS OF THE THIRD HARMONIC OF A VACUUM\\
RESPONSE IN A SUBCRITICAL LASER FIELD}

\author{\firstname{V.~V.}~\surname{Dmitriev}}
\email{dmitrievvv@info.sgu.ru}
\affiliation{%
Saratov State University, Saratov, RU, 410026, Russia
}%
\author{\firstname{S.~A.}~\surname{Smolyansky}}
\email{smol@info.sgu.ru}
\affiliation{%
Saratov State University, Saratov, RU, 410026, Russia
}%
\affiliation{%
Tomsk State University, Tomsk, RU, 634050, Russia
}%

\author{\firstname{V.~A.}~\surname{Tseryupa}}
\affiliation{%
Saratov State University, Saratov, RU, 410026, Russia
}%


\begin{abstract}
We investigate a nonlinear response of the physical vacuum by the example of the third harmonic in radiation of the electron-positron-photon plasma exciting by a strong time dependent electric field at the focus spot of counterpropagating laser pulses. The investigation was developed within the framework of the Bogoliubov-Born-Green-Kirkwood-Yvon kinetic theory nonperturbatively describing vacuum creation of the quasiparticle electron-positron plasma (EPP) and different channels of its interaction with the photon subsystem in the single-photon approximation that are opened in the presence of a strong quasiclassical field. We investigate the radiation of EPP in the annihilation channel on the basis of the well known approximations and compare the obtained result with the quasiclassical radiation of the plasma oscillations. In the high-frequency range for large adiabacity parameter the quantum radiation dominates in comparison with the quasiclassical one.
\end{abstract}

\maketitle

\section{Statement of the problem}

Theoretical and experimental investigation of nonlinear effects in
the   strong   field  QED is one of the central problems of modern physics. In  this
communication, we present  some  new results of investigation of the kinetic
theory of the electromagnetic radiation of the electron-positron
plasma  (EPP) generated from vacuum under action of a strong spatially
homogeneous time  dependent electric field. This kinetic
theory   has  a nonperturbative   foundation  for  describing the interaction  of EPP
with  a  strong quasiclassical field (recent reviews \cite{1,2})
whereas  the  interaction with  a quantized electromagnetic field is
considered  in  the  second order of the perturbation theory
\cite{3,4}. Single-photon channels are closed under ordinary conditions,
however, the interaction of EPP with  a strong external field can open them
\cite{5}.

Let the strong quasiclassical electric field is given as $A^\mu(t)=(0, 0, 0, A^3(t)=A(t))$. For this case the complete system of kinetic equations (KEs) for EPP (with distribution function $f({\bf p},t)$) and photons (with distribution function $F({\bf k},t)$) in the strong field QED in the single-photon approximation can be written in the form \cite{3,4}
\begin{eqnarray}\label{eq1_smol}
\dot{f}({\bf p},t) &=& I({\bf p},t)+C^{(e)}({\bf p},t)+C^{(\gamma)}({\bf p},t),\\ \label{eq2_smol}
\dot{F}({\bf k},t) &=& S^{(e)}({\bf k},t)+S^{(\gamma)}({\bf k},t),
\end{eqnarray}
where $I({\bf p},t)$ is a source term describing vacuum production of EPP \cite{1,2},
\begin{equation}\label{eq3_smol}
I({\bf p},t)=\frac{1}{2}\lambda({\bf p},t)\int_{t_0}^{t}dt' \lambda({\bf p},t')\left[1-2 f({\bf p},t')\right]\cos \theta (t,t'),
\end{equation}
where
\begin{equation}\label{eq4_smol}
\lambda({\bf p},t)=\frac{e E(t) \varepsilon_\perp}{\omega^2 ({\bf p},t)},\ \omega^2 ({\bf p},t)=\sqrt{\varepsilon^2_\perp+P^2},\ \varepsilon_\perp=\sqrt{p^2_\perp+m^2},\ P=p_\parallel-eA(t), 
\end{equation}
\begin{equation}\label{eq5_smol}
\theta (t,t')=2\int_{t'}^{t} d \tau \ \omega({\bf p},\tau).
\end{equation}
In KEs (\ref{eq1_smol}), (\ref{eq2_smol}) the collision integrals (CIs) $C^{(e,\gamma)}$ and $S^{(e,\gamma)}$ are quadratic functionals with respect to $f$ and linear with respect to $F$.
KEs (\ref{eq1_smol}), (\ref{eq2_smol}) allow us to study a wide range of phenomena: single-photon annihilation $(C^{(e)})$ and photoproduction $(C^{(\gamma)})$ as well as emission $(S^{(e)})$ and absorption $(S^{(\gamma)})$.

To diagnose EPP, it is necessary to take into account the second mechanism of EPP radiation, based on plasma oscillations of EPP. It brings to the quasiclassical radiation of EPP, in contrast to the quantum character of the electromagnetic field described by KE (\ref{eq2_smol}). This situation leads to the problem of identifying radiation mechanisms from experimental results. The present paper summarizes preliminary investigations of this problem within the framework of some simplest model approximations.

Thus, the starting KE for investigation of the quantum radiation of EPP is KE (\ref{eq2_smol}) with the CI in the annihilation channel \cite{4}
\begin{equation}\label{eq6_smol}
S^{(e)}({\bf k},t)=\int \frac{d^3p_1}{(2\pi)^3}\int \frac{d^3p_2}{(2\pi)^3}
\int^{t}dt' K^{(e)}({\bf p}_1,{\bf p}_2,{\bf k},t,t')f({\bf p}_1,t')f({\bf p}_2,t'),
\end{equation}
with the kernel
\begin{equation}\label{eq7_smol}
K^{(e)}({\bf p}_1,{\bf p}_2,{\bf k},t,t')=\frac{(2\pi)^3 e^2 \delta ({\bf p}_1-{\bf p}_2-{\bf k})}{2k\omega({\bf p}_1,t)\omega({\bf p}_2,t)}\Delta^{(+)}({\bf p}_1,{\bf p}_2,{\bf k},t)\cos\theta^{(+)}({\bf p}_1,{\bf p}_2,{\bf k},t,t'),
\end{equation}
\begin{equation}\label{eq8_smol}
\Delta^{(+)}({\bf p}_1,{\bf p}_2,{\bf k};t)=2(e^r({\bf k})P_1) (e^r(-{\bf k})P_2)-[(P_1 P_2)+m^2](e^r({\bf k}) e^r(-{\bf k}))
\end{equation}
and the phase is
\begin{equation}\label{eq11}
\theta^{(+)}({\bf p}_1,{\bf p}_2,{\bf k},t,t')=\int_{t'}^{t}d\tau[\omega({\bf p}_1,\tau)+\omega({\bf p}_2,\tau)-k].
\end{equation}
In the CI (\ref{eq6_smol}) we additionally neglect the back reaction from the quantum field.

Subsequent calculations are based on two reliable approximations.

1. The distribution function $f({\bf p},t)$ of EPP in CI (\ref{eq6_smol}) is limited by the self-consistent field approximation, so $\dot{f}=I$, where the source term $I({\bf p},t)$ is defined by Eq. (\ref{eq3_smol}). Additionally, the low density approximation is introduced in this KE based on the substitution $1-2f\to1$ in the source term  (\ref{eq5_smol}), so
\begin{equation}\label{eq10_smol}
\dot{f}({\bf p},t)=I({\bf p},t)\approx\frac{1}{2}\lambda({\bf p},t) \int_{-\infty}^t dt' \lambda({\bf p},t')\cos\theta(t,t').
\end{equation}
From here it follows \cite{6}
\begin{equation}\label{eq11_smol}
f({\bf p},t)=\frac{1}{2}\int^t dt' \lambda({\bf p},t') \int^{t'} dt'' \lambda({\bf p},t'')\cos\theta(t',t'').
\end{equation}
 Effectiveness of this approximation in the subcritical fields $E\lesssim E_c=m^2/e$ has been verified in the work \cite{7}.

2. In addition, the effective electromagnetic mass approximation is introduced \cite{8}, based on substitution into the quasienergy (\ref{eq4_smol}),
\begin{equation}\label{eq12_smol}
\omega({\bf p},t)\to\omega^*(p)=\sqrt{m^{*2}+p^2},
\end{equation}
\begin{equation}\label{eq13_smol}
m^{*2}=m^2+\frac{1}{2}(eE_0/\Omega)^2=m^2+\frac{1}{2}m^2\gamma^{-2},
\end{equation}
where $\gamma=\Omega E_c/(mE_0)$ is the Keldysh parameter. Here it is assumed that the external field is harmonic,
\begin{equation}\label{eq14_smol}
A(t)=-\frac{E_0}{\Omega}\sin\Omega t, \ \ \ E(t)=E_0\cos\Omega t.
\end{equation}
Now we can discuss some consequences of the inserted approximations in the CI (\ref{eq6_smol}).

\section{Estimations of the third harmonic}

The momentum conservation law ${\bf p}_1-{\bf p}_2-{\bf k}=0$ in the single-photon model is contained in the kernel (\ref{eq6_smol}) whereas the energy conservation law in the annihilation channel appears as an argument of the $\delta$-function that occurs as a result of calculating the time integral in CI (\ref{eq6_smol}) in the asymptotic limit $t\to\infty$ after bringing in of the introduced approximations,
\begin{equation}\label{eq15_smol}
\sigma({\bf p}, {\bf k} \ | \ \Omega)=\omega^*({\bf p})+\omega^*({\bf p}+{\bf k})-k-n\Omega=0.
\end{equation}
Here the integer $n=0, 1, ...,8$ is a consequence of the polynomial structures with respect to the potential $A(t)$ (\ref{eq14_smol}) of the kernel (\ref{eq7_smol}), (\ref{eq8_smol}) and the distribution function (\ref{eq11_smol}) written in the effective mass approximation (\ref{eq12_smol}), (\ref{eq13_smol}). The conservation law (\ref{eq15_smol}) in the case when the external field is switching off ($\Omega\to 0$ and $\omega^*({\bf p})\to\omega_0({\bf p})=\sqrt{m^2+p^2}$) transforms in $\omega_0({\bf p})+\omega_0({\bf p}+{\bf k})-k=0$ which is incompatible with the momentum conservation law. The switching on of an external field can activate this channel. The condition of activation of this channel follows from Eq. (\ref{eq15_smol}): $2m^*=n_c\Omega$, where $n_c$ is a minimal number of photons from the photon reservoir of the external field (\ref{eq14_smol}). In the case of the polynomial nonlinearity the minimal presence of the external field corresponds to the maximal degree of potential $A(t)$ in the integrand CI (\ref{eq6_smol}) (in the framework of the introduced approximations). In the case under consideration $n_c=8$ and hence $\Omega_c=m^*/4$ and $\sigma=0$ according to Eq. (\ref{eq15_smol}).

Now we can find the roots of Eq. (\ref{eq15_smol}) with $\Omega=\Omega_c$:
\begin{equation}\label{eq16_smol}
p^{(1,2)}=k\cos\alpha \frac{2m^*(k+m^*)}{(k+2m^*)^2-k^2\cos^2\alpha}\pm\sqrt{\Delta^*},
\end{equation}
\begin{equation}\label{eq17_smol}
\Delta^*=
k^2\cos^2\alpha \left[ 1+\frac{m^{*2}}{(k+2m^*)^2} \right]
+4m^{*2}\left( 1-\frac{m^*}{k+2m^*} \right)^2-m^{*2}\geq 0,
\end{equation}
where $\alpha$ is the angle between vectors ${\bf k}$ and ${\bf p}$.

Let us consider the quantum radiation in a transverse direction relative to the external field, ${\bf k}{\bf E}=0$ (it can be shown from the analysis of function (\ref{eq8_smol}) that the radiation is absent in a longitudinal direction).

$\delta(\sigma)$ function with the roots (\ref{eq16_smol}), (\ref{eq17_smol}) allows us to calculate the integral over $p=|{\bf p}|$ in CI (\ref{eq6_smol}) and analyse the residual angular integrals in all frequency range of radiation. Below we will be limited by a rough estimation of the power of the quantum radiation on a frequency of the third harmonics of the external field $(k=3\Omega)$ only and will compare it with the power of the quasiclassical radiation generated by inner plasma oscillations. The parameter of adiabaticity in such a high-frequency limit is large, $\gamma\gg 1$ and hence $m^*\approx m$. Then it can be shown that the integrand in CI (\ref{eq6_smol}) is regular everywhere and can be substituted by the constant at $\alpha=0$ in Eqs. (\ref{eq16_smol}), (\ref{eq17_smol}).

These approximations bring to the following rough estimation for a photon production rate ($\alpha=e^2/4\pi$):
\begin{equation}\label{eq18_smol}
\frac{dF_\perp({\bf k},t)}{dt}\Bigg|_{t\to\infty}=S^{(e)}_\perp({\bf k}) \sim \alpha m^* \left(\frac{m}{\gamma m^*}\right)^8.
\end{equation}
The simplest situation corresponds to a large parameter of adiabaticity $\gamma\gg 1$ ($E_0\lesssim E_c$ and large frequency). Then $m^*\approx m$ and
\begin{equation}\label{eq19_smol}
S^{(e)}_\perp({\bf k}) \sim \alpha m/\gamma^8.
\end{equation}

It would be interesting to compare this result with the quasiclassical radiation generated by plasma oscillations \cite{9}. Below we adopt the main result of this work for actual parameters of the external field ($E_0=0.1E_c$, $\Omega=m/4$) introduced above. Fig. 1 allows us to estimate effectiveness of the external field transformation, in particular, in the third harmonic of nonlinear plasma oscillations.

Let us introduce the transformation coefficient of the external field (\ref{eq14_smol}) with the time-average energy density $E^2_0/16\pi$ into the radiation energy (\ref{eq19_smol}) per space cell of volume $(2\pi/k)^3$
\begin{equation}\label{eq20_smol}
R_{quant}=\frac{8\alpha^2\Omega}{\pi^2\gamma^6m}\sim 10^{-10}.
\end{equation}

External field in the work \cite{9} was defined as an impulse field, $E(t)=E_0\cos\Omega t \cdot$ $\exp[-t^2/(2\tau^2)]$. In this case the transformation coefficient of the external field into quasiclassical radiation is introduced as the ratio between the spectral energy densities of the third harmonic of internal plasma field $w_{in}(3\Omega)$ and the external field one $w_{ex}(\Omega)$,
\begin{equation}\label{eq21_smol}
R_{q.c.}=\frac{w_{in}(3\Omega)}{w_{ex}(\Omega)}=\frac{|E_{in}(3\Omega)|^2}{|E_{ex}(\Omega)|^2}\sim 10^{-14}.
\end{equation}

Comparison of the estimations (\ref{eq20_smol}), (\ref{eq21_smol}) shows considerable dominance of the quantum radiation over quasiclassical one.

The obtained rough estimations open up  possibilities for a more detailed investigation both annihilation channel of radiation from EPP and other channels in the single-photon approximation within the framework of the kinetic theory of the electron-positron-photon plasma generated in strong electromagnetic fields from the vacuum  \cite{3,4}.

The fulfilled investigation shows also that the quantum mechanism of the high-energy photon radiation can play significant role in strong field evolution of EPP. It offers perspective for research of the exhaustion effect of EPP in strong fields with $E_0\sim E_c$ \cite{10} on the basis of the self-consistent kinetic theory of the electron-positron-photon plasma \cite{3,4}.

\newpage

\begin{figure}
\setcaptionmargin{5mm}
\onelinecaptionsfalse
\includegraphics{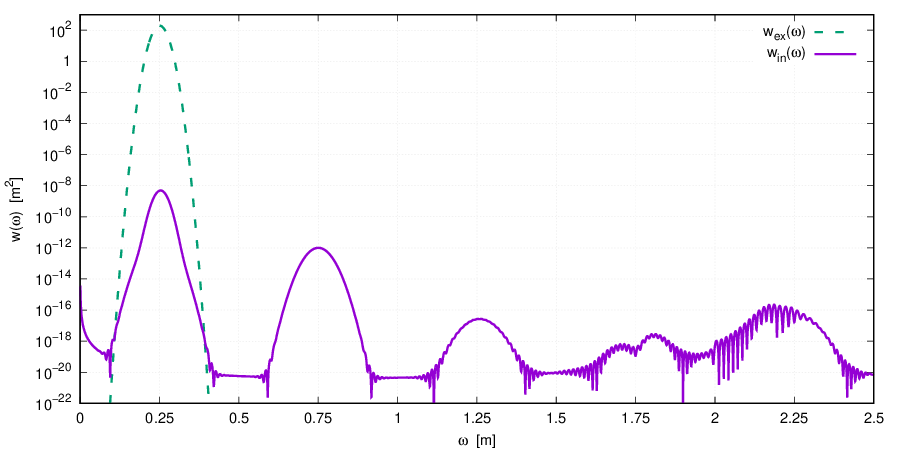}
\captionstyle{normal}
\caption{Regularized spectral energy density of the quasiclassical radiation $w_{in}(\omega)$ and the impulse external field $w_{ex}(\omega)$.}
\end{figure}

\end{document}